\documentclass[aps,prc,preprint,groupedaddress,showpacs]{revtex4}

\usepackage{graphicx}

\begin{document}
\title{Evidence for strong refraction of $^3$He in an 
alpha-particle  condensate }

\author{S. Ohkubo$^1$   and   Y. Hirabayashi$^2$ }

\affiliation{$^1$Department of Applied Science and Environment,
Kochi Women's University, Kochi 780-8515, Japan  }

\affiliation{$^2$Information Initiative Center,
Hokkaido University, Sapporo 060-0811, Japan}

\date{\today}


\begin{abstract}
\par
We have analyzed $^{3}$He scattering from $^{12}$C at 34.7 and  72 MeV
in a coupled channel method  with a double folding 
potential derived from  the 
precise wave functions for the ground 0$^+$ state and  $0_2^+$ (7.65 MeV) 
 Hoyle state, which has been suggested to be an  $\alpha$ particle 
condensate.  
It is found that strong refraction of $^3$He in the Hoyle state
can be clearly seen in the experimental angular distribution at {\it low}
 incident energy  region as an Airy minimum of the 
{\it pre-rainbow oscillations}.

\end{abstract}


\maketitle

\par
  Bose-Einstein condensation (BEC) has been well established in a dilute gas
\cite{Leggett2001}. The remarkable properties of superconductivity and 
superfluidity  in both $^3$He  and $^4$He are related to BEC. In hadronic systems
of strong interaction BEC has been paid much attention in pion condensation and 
kaon condensation. To the best of our
  knowledge, 
 refractive effect of
     BEC  states in the scattering  has  been rarely investigated. 
In this paper we show that  strong refractive effect is observed in 
the Hoyle state of 
$^{12}$C, which has been suggested to be  an
 $\alpha$ particle condensate  in nuclei.  

  Pioneering work of Hokkaido group by  Uegaki {\it et al.} 
 \cite{Uegaki1979} and Kamimura and Fukushima  \cite{Kamimura1981}  
showed in the microscopic cluster model that the $0_2^+$ state of $^{12}$C,
 the Hoyle state,  
  has a loosely coupled three $\alpha$ cluster structure with an 
$\alpha$$\otimes$$^8$Be configuration. 
 Recently it has been  shown that the wave functions of Uegaki
 {\it et al.} \cite{Uegaki1979} 
and Kamimura and Fukushima \cite{Kamimura1981}  are  almost completely equivalent to the wave
function that the three $\alpha$ particles are sitting in the lowest 0s state
  like a dilute gas and speculated  that the Hoyle state is a  
 Bose-Einstein condensate of  three $\alpha$ particles
  \cite{Tohsaki2001,Funaki2003,Roepke1998}. 

 In a BEC of atomic gas the system is magnetically trapped. 
 On the other hand, as the nucleus is a 
self-binding finite system due to strong nuclear interaction, there is no external
field to trap the system. 
Although the BEC of atomic gas appears at near {\it zero temperature}, the gaseous state
 of weakly interacting $\alpha$ particles like the Hoyle state
appears at a {\it highly excited energy}  near the  threshold. 
If such a dilute state  due to BEC exists,  typical macroscopic 
 physical quantities peculiar to it would  exist. The
 huge radius  of its state may be one of them.  To measure 
such a huge  radius of the excited state is very challenging. However,
no such an experiment has been reported. 
Recently Kokalova {\it et al.} \cite{Kokalova2006} proposed a new
experimental  way of testing BEC of $\alpha$
particles in nuclei by directly observing the enhancement of $\alpha$
particle emission  and the multiplicity  partition of the possible
 emitted
$\alpha$ particles. 

In this paper we show that the strong refractive effect of incident $^3$He in 
 the  Hoyle state of 
$^{12}$C  can be seen in  the {\it  pre-rainbow oscillations}  
in the {\it low} incident  energy region  where there is a pocket in the effective potential
 (nuclear plus Coulomb plus centrifugal). 
Usually refractive effect has been discussed as nuclear rainbow scattering 
in the  {\it high} incident 
 energy region where the analogy between the meteorological rainbow and the 
 nuclear rainbow can be discussed based on the classical
  deflection function
   \cite{Ford1959,Nussenzveig1977}.
We  discussed the relation between the nuclear rainbow and BEC in $\alpha$ particle scattering from $^{12}$C($0^+_2$) at {\it high} incident  
energies 
in a previous paper \cite{Ohkubo2004} because experimental data  
 were only  available  above $E_\alpha$=139 MeV.
However, it is expected  that the  refractive effect becomes much larger 
and can be seen clearly at {\it low}  incident energies. 
In fact, in optics \cite{Born1999} refractive index $n$ is  related
 to the optical potential $V$ as follows:
 \begin{equation}
 n(r) = \sqrt{1-\frac{V(r)}{E_{c.m.}}} 
\end{equation} 
\noindent  Also for     $\alpha$ particle scattering from $^{12}$C
there is a  well-known long-standing difficulty that a  
global  potential for the $\alpha$+$^{12}$C 
system has not been known in the  {\it low} incident   energy  region \cite{Hirabayashi2002}.

  The pre-rainbow structure  has been mostly studied in 
elastic scattering \cite{Michel2000,Michel2002}. There had been 
no systematic  theoretical and experimental 
studies of  the pre-rainbow oscillations in {\it inelastic} scattering. 
However, recently  it has been shown that 
  inelastic  pre-rainbow Airy structure can also be  understood in 
a  way similar to  elastic scattering 
\cite{Michel2004B}.
 This suggests that  inelastic pre-rainbow  oscillation may also be useful for
 the study of   the nuclear properties of the excited states of the target nucleus 
  because the internal region of the interaction potential 
can be  well determined even for  inelastic channels.

  Fortunately the pre-rainbow Airy structure in the scattering from the
 Hoyle state   was measured in the low incident  energy region $E_L$=34.7 MeV 
many years ago 
in $^3$He  scattering \cite{Fujisawa1973} as well as the high energy  scattering at   72 MeV 
 \cite{Demyanova1992} where the falloff of the cross sections 
is seen in the experimental data.
However, unfortunately these inelastic scattering data from the  Hoyle   state have 
 been forgotten
and have  never been studied from the theoretical point of view.
 Also elastic  $^3$He scattering  from $^{12}$C 
has not been
studied systematically compared with  $\alpha$+$^{12}$C
 scattering.
Most of the analyses of elastic $^3$He scattering from nuclei 
 have been done with  a conventional Woods-Saxon potential \cite{Trost1987}. 
  Khallaf {\it et al.} \cite{Khallaf1997,Khallaf2003} have  analyzed
  $^3$He+$^{12}$C elastic scattering  with a folding
 potential.
 
 We study  elastic and inelastic $^3$He+$^{12}$C scattering
  in the microscopic coupled channel method by taking into 
 account simultaneously the  0$^+_1$ (0.0 MeV), $2^+$ (4.44 MeV),
 0$^+_2$ (7.65 MeV),  and 3$^-$ (9.63 MeV) states of $^{12}$C.
The diagonal and coupling potentials for the $^3$He+$^{12}$C
 system are calculated by the double folding  model:

\begin{equation}
V_{ij}({\bf R}) =
\int \rho_{00}^{\rm (^3He)} ({\bf r}_{1})\;
     \rho_{ij}^{\rm (^{12}C)} ({\bf r}_{2})\;
v_{\rm NN} (E,\rho,{\bf r}_{1} + {\bf R} - {\bf r}_{2})\;
{\rm d}{\bf r}_{1} {\rm d}{\bf r}_{2}  \; ,
\end{equation}

\noindent
where $\rho_{00}^{\rm{ (^3He)}} ({\bf r})$ is the ground 
state density
of  $^3$He taken from Ref.\cite{Cook1981}, while $v_{\rm NN}$ denotes
 the density-dependent M3Y effective interaction (DDM3Y) \cite{Kobos1984}.
$\rho_{ij}^{\rm (^{12}C)} ({\bf r})$ represents the diagonal 
($i=j$) or transition ($i\neq j$) nucleon density of $^{12}$C
calculated in the resonating group method by 
Kamimura {\it et al.} \cite{Kamimura1981}.
The folding potential
is very sensitive to the  wave functions used, which serves
 as a good test of the validity of the 
 wave function \cite{Khoa1995}. 
 This wave function  for the
   Hoyle state  is  almost completely equivalent 
 to
the  Bose-Einstein condensate wave function \cite{Funaki2003}.
 In the analysis  we  introduce the normalization factor 
 $N_R$ for 
 the real part of the potential and phenomenological
 imaginary potentials with a  Wood-Saxon form factor  (volume absorption) 
and a derivative of the  Wood-Saxon form factor (surface absorption) 
  for each channel. 

  \begin{figure}[htb]
\includegraphics[width=14cm]{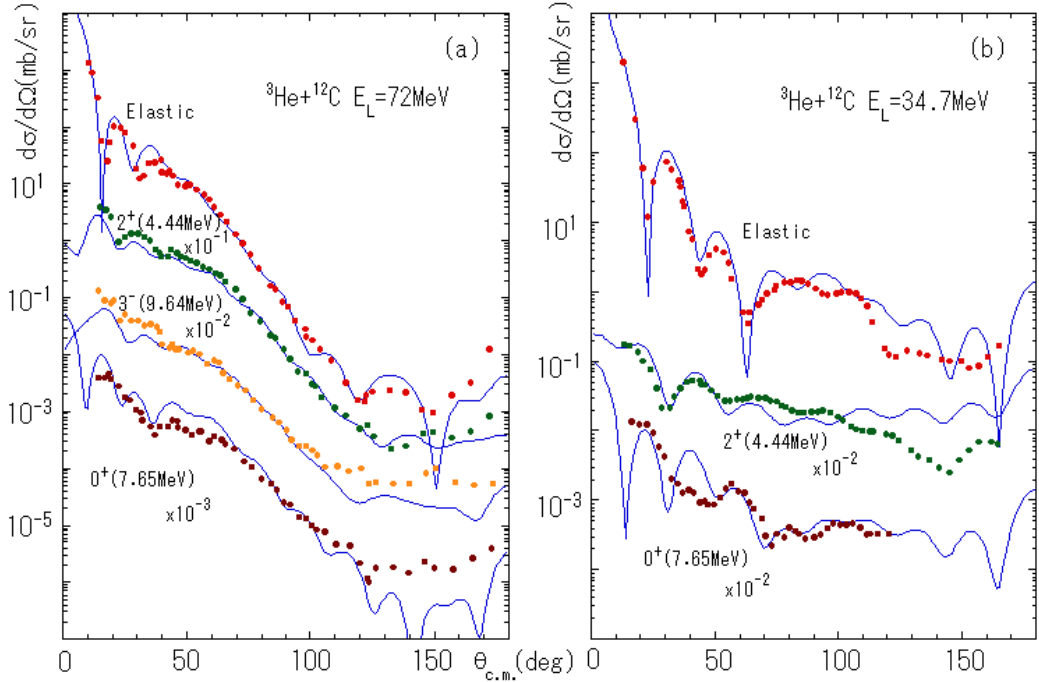}
\caption{\label{fig.1} {(Color online) The calculated elastic and inelastic 
 angular distributions (solid lines) for the $2^+$ (4.44MeV),  
$0^+_2$ (7.65 MeV)  and $3^-$ (9.63 MeV) states of  $^{12}$C in 
 $^3$He+$^{12}$C scattering at $E_L$=72  and  34.7 MeV      are  compared with
   the experimental   data (points)   \cite{Fujisawa1973,Demyanova1992}.
 }
}
\end{figure}

   In Fig.~1(a)    calculated  angular
 distributions  are shown in comparison with the 
 experimental  data of   elastic and inelastic  $^3$He
  scattering from   $^{12}$C at $E_L$=72 MeV.
       The characteristic features of the  falloff of the cross sections
 beyond the rainbow angle in  the
  experimental  angular distributions for the shell-like   ground, 
 $2^+$, 3$^-$  states, and the 
  0$^+_2$   state with the well-developed $\alpha$-cluster 
structure are simultaneously well  reproduced. 
It is noted that for the ground  and the  $2^+$  
 states  the 
agreement of the calculations with the data is fairy good up to  
large angles. 
      In Fig.~1(b) the same comparison is  shown for  the experimental
 data at 34.7 MeV  \cite{Fujisawa1973}. In this energy region the 
 falloff of the cross sections in the angular distributions 
characteristic to  rainbow scattering is  no more seen in the
 experimental data. The  broad bumps in the intermediate
angular region of the angular distributions in elastic and inelastic 
scattering are reproduced by the calculations.
The discrepancy  between the experimental data and the
 calculation  seen only
 for  the  0$^+_2$ state
  at forward angles where the nearside contributuion increases (Fig.~2(b))
  may be mostly due to the truncation of the explicit coupling to the 
higher
 excited states.  In fact, for example,  most
 of the imaginary potential for the  0$^+_2$  comes from the  coupling 
to  the 2$^+_2$ state, which has a well-developed $\alpha$-cluster
 structure with almost the same configuration as the 0$^+_2$ state. 
We confirmed that the present calculation also reproduces the experimental 
angular distribution at $E_L$=119 MeV of Hyakutake {\it et al.}
 \cite{Hyakutake1980} very well.

   To see the refractive effect  in  Fig.~2 the calculated angular
 distributions are decomposed 
  into farside and nearside components following the
Fuller's prescription \cite{Fuller1975}. 
   At $E_L$=72 MeV the first Airy minimum $A1$  appears at 30$^\circ$
for elastic scattering and 35$^\circ$ for the $0^+_2$ state. For elastic scattering
a clear minimum is not seen in the angular distribution of the 
farside cross sections and the  Airy  minimum
in the experimental data at 30$^\circ$ is obscured  by the interference between
 the farside and nearside amplitudes. 
 On the other hand,
the $A1$ minimum for  the 0$^+_2$ state  is clearly seen in the farside cross sections
 because the minimum is shifted to a larger angle where  the nearside 
  contribution is  much smaller.
 The situation is more clearly seen in the Airy structure at {\it low} incident  energy region
where there is a pocket in the effective potential and no typical rainbow 
falloff of the dark side appears.  At $E_L$= 34.7 MeV in  Fig.~2(b)  the Airy minimum $A1$
 appears at 60$^\circ$ for elastic scattering and 75$^\circ$ for the 0$^+_2$ state.
 The latter is much shifted to a larger angle and the Airy minimum is not at all 
obscured 
 by the nearside contributions, which decreases rapidly as the scattering angle 
increases.
For the 0$^+_2$ state, in the  wider range of angles the nearside contributions 
are 
much smaller than the farside contributions compared with the elastic scattering
 case.
Thus the difference of the refraction between the ground state and the 0$^+_2$ state is much
more clearly seen at 34.7 MeV than at 72 MeV.  The Airy minimum for the 0$^+_2$ state
is three times more shifted to a larger angle than the case of $E_L$=72 MeV.
 The absorption
is incomplete for the 0$^+_2$ state and a more beautiful pre-rainbow Airy oscillation  is 
seen than the elastic scattering. Thus we can say that refractive effect of the 0$^+_2$
 state is more clearly seen  in the pre-rainbow  structure at {\it low} incident energy
  region than at high incident energy region
 where  a typical nuclear  rainbow appears.

\begin{figure}[tb]
\includegraphics[width=14cm]{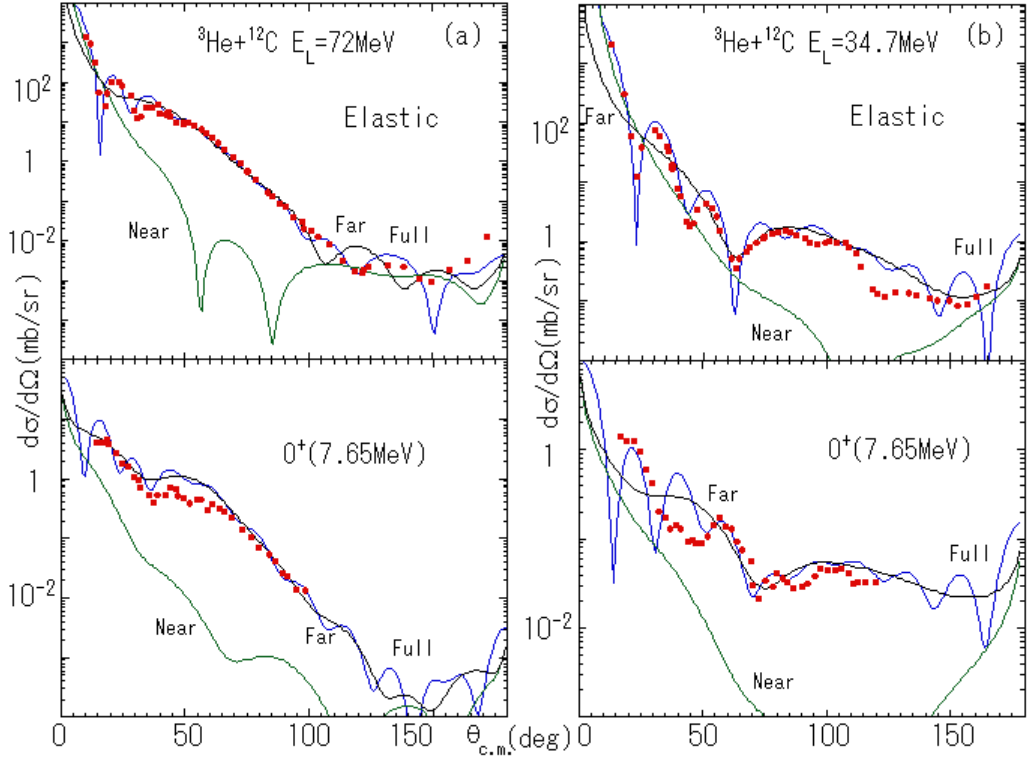}
\caption{\label{fig.2} {(Color online)  
The angular distributions for the ground  and  $0^+_2$ (7.65 MeV) states 
of $^{12}$C in 
     $^3$He+$^{12}$C scattering at $E_L$=72  and 34.7 MeV MeV   are  decomposed into 
 farside (dashed lines) and  nearside ( dotted lines) contributions. 
 }
}
\end{figure}

\begin{table}[th]
\begin{center}
\caption{The  volume integral per nucleon
pair $J_V$ , root mean square radius $<R^2>^{1/2}$, of the
folding potential,  and
  the  parameters of the imaginary potentials in the conventional
notation.
The normalization factor $N_R$ is fixed to 1.28.
 }
\begin{tabular}{cccccccccc}
 \hline
  \hline
  $E_L$ & $J^\pi$ &$J_V$ &  $<R^2>^{1/2}$ & $W_V$ & $R_V$ & $a_V$ & $W_S$ &
$R_S$ & $a_S$ \\
  (MeV) &   &     (MeV fm$^3$) & (fm)&(MeV) & (fm) & (fm)& (MeV) & (fm) &
(fm) \\
 \hline
   72 & $0_1^+$  &  408 &  3.576 & 5.0  & 5.6 & 0.60 & 4.0  & 2.6  & 0.20
\\
      & $2^+$    &  403 &  3.562 & 8.0  & 5.0 & 0.10 & 7.0  & 2.7  & 0.30
\\
      & $3^-$    &  461 &  3.827 & 9.0  & 4.9 & 0.10 & 9.0  & 2.4  & 0.10
\\
      & $0_2^+$  &  529 &  4.385 & 18.0 & 4.5 & 0.20 & 12.0 & 2.6  & 0.50
\\
 34.7 & $0_1^+$  &  445 &  3.574 & 6.0  & 4.8 & 0.50 & 6.0  & 2.6  & 0.50
\\
      & $2^+$    &  440 &  3.559 & 4.0  & 4.3 & 0.30 & 6.0  & 2.4  & 0.40
\\
      & $3^-$    &  506 &  3.829 & 9.0  & 4.9 & 0.40 & 9.0  & 2.6  & 0.50
\\
      & $0_2^+$  &  586 &  4.385 & 19.0 & 5.4 & 0.60 & 14.0 & 2.8  & 0.60
\\

 \hline
 \hline
\end{tabular}
\end{center}
\end{table}

In Table I the properties of the real folding potential and 
 imaginary potential parameters used are given.
 The   obtained volume integral per nucleon  pair of the potential
 $J_V$=365  MeVfm$^3$ for elastic scattering at $E_L$=119 MeV is consistent with 
 352  MeVfm$^3$  of the unique potential of Hyakutake {\it et al.} \cite{Hyakutake1980}. 
 The obtained $J_V$=408 MeVfm$^3$  at $E_L$=72 MeV for elastic scattering is closer to a deeper potential
 set  $A4$ ( $J_V$=437 MeVfm$^3$)  than a shallower set set A2 
( $J_V$=275 MeVfm$^3$) of Dem'yanova  {\it et al.} \cite{Demyanova1992}.
The volume integral for elastic scattering increases as energy decreases and 
becomes  445 MeVfm$^3$ at $E_L$=34.7 MeV. 
The energy dependence of the present double folding potential for the $^3$He+$^{12}$C
 system is consistent  with a systematic study in Ref.~\cite{Trost1987}.
 The volume integral for $^3$He scattering is deeper  than that
 for $\alpha$-particle scattering although   its  energy evolution 
 is similar. 
   We compare the refractive effect for the ground state and the 
   0$^+_2$ state at  $E_L$=34.7 MeV.
    The obtained  $J_V$=583MeVfm$^3$  for   the  0$^+_2$ state
    is much larger than $J_V$=445 MeVfm$^3$ for the ground state.     
    The calculated root mean square radius of the 
   real potential is 4.385 fm for the 0$^+_2$ state and 3.574 fm  for the ground 
state. This  shows that the refractive effect is extremely stronger for the 0$^+_2$ state
   than the ground state, and that the potential, that is the lens, 
 for the 0$^+_2$ state is much  more extended than that for the ground state in
 agreement with the dilute     distribution of the density  of the 0$^+_2$ state.

  From the experimental data itself of the Airy minimum of the 0$^+_2$ state
we can know that the density distribution of the 0$^+_2$ state is far more
extended and dilute than the ground state.
For elastic scattering the rainbow angle $\theta_N$ can be given analytically
if the Wood-Saxon form factor is assumed for the real part of the 
optical potential \cite{Knoll1976}:
\begin{equation}
\theta_N  \approx \mid V_C-0.56V  (\frac{R}{a})^{\frac{1}{2}} \mid /E_{c.m.}
\end{equation}
\noindent where $V_C=Z_1Z_2e^2/R$ and $V$, $R$ and  $a$  are the  strength,
 radius and diffuseness parameters of the potential, respectively.
This means that  the  larger $\theta_N$ is, the  deeper the potential 
strength is (or the larger the radius is). This also means that the more 
 the first Airy
minimum is shifted to a larger angle, the deeper the potential becomes (or 
the larger the radius of the potential becomes). Based on the similarity of 
 the systematic energy evolution in a wide range of incident energies
of the Airy structure between elastic scattering and  inelastic
 scattering \cite{Michel2004B}, the above discussion between the 
position of the first  Airy minimum of the pre-rainbow oscillations and the depth (and the radius) of the potential 
will hold qualitatively in the present case. Because the Airy minimum of the 
pre-rainbow oscillations for the
 0$^+_2$ state
appears at a larger angle as seen in Fig.~2(b),  the radius parameter $R$ 
 of the corresponding potential is far larger than
 that for the ground state considering that the depth $V$ of the potential 
for the $0^+_2$ is smaller
 than that for the ground state. Therefore the shift of the 
 angular position  of the Airy minimum of the  pre-rainbow oscillations for the 0$^+_2$ state from that for 
 the ground may be used to  measure  the size of the lens.

 The present approach   to know the size of the lens of  the excited state
qualitatively,
namely how dilute is the excited state can  be applied to the $n\alpha$-particle
 states    of $4N$-nuclei  near the threshold such as   the  four $\alpha$-particle
 state  in    $^{16}$O and ten $\alpha$-particle state in $^{40}$Ca by using
 $\alpha$ particle, $^3$He and $^{16}$O  as a projectile, for which absorption is 
incomplete.
For non-$4N$ nuclei the  ${\frac{3}{2}}^-$ state at  8.56 MeV 
 in $^{11}$B analogue to the 0$^+_2$ state of $^{12}$C  is considered to have
 a dilute density distribution  \cite{Kawabata2006}.
Also   it has been  suggested that  analogue states  appear in neutron rich nuclei, for example, 
 the  $\frac{1}{2}^-$(8.86 MeV)  state in $^{13}$C 
\cite{Milin2002} , the  $0^+$(9.746 MeV) state in 
$^{14}$C \cite{Oertzen2004} and the $0^+$( $\sim$29 MeV) state in $^{16}$C \cite{Itagaki2001}, which are  considered 
to have one, two and four   
 additional neutrons to the  0$^+_2$ state of  $^{12}$C. 
  It is interesting to observe systematically how the  pre-rainbow Airy minimum 
 in inelastic scattering    at low incident  energy region is shifted as  additional
 neutrons are added  (removed)  to (from) $^{12}$C.

To summarize, it is found that strong refraction of $^3$He in 
   the 0$_2^+$ (7.65 MeV) Hoyle state of  $^{12}$C,
    which has  been suggested to be an  $\alpha$ particle 
condensate,  
can be clearly seen in the experimental angular distribution at  {\it low}
 incident energy  region where there is a pocket in the
effective potential as an Airy minimum of the 
{\it pre-rainbow oscillations}.  Because  of this strong refraction,
 the Airy minimum
is    shifted to
a larger angle considerably compared with that of  the normal ground state,
it is clearly observed being not obscured by the nearside contributions.
 The present finding may also hold for  
$\alpha$ particle condensate in heavier nuclei like $^{40}$Ca 
and  excited states with dilute density distribution.

  One of the authors (S.O.)   has been supported by a
 Grant-in-aid for Scientific Research
 of the Japan Society for Promotion of Science (No. 16540265)
and the Yukawa Institute for Theoretical Physics.

\end{document}